# Questões à "Interpretação da Consciência Processual" da Mecânica Quântica*

Raoni Arroyo[1]
Lauro de Matos Nunes Filho[2]
Frederik Moreira dos Santos[3]

> "Between the potency
> And the existence
> […]
> Falls the Shadow."
> ______________________________
> (T. S. Eliot, "The Hollow Men")

## 1 Introdução

A *interpretação da consciência processual* (ICP), conforme desenvolvida em Arroyo (2024, cap. 5) e Arroyo; Nunes Filho; Moreira dos Santos (2024), propõe uma ontologia de processos como solução ao problema da medição. Este artigo apresenta a interpretação padrão levada às últimas consequências ontológicas (§2); apresenta a ICP (§3); propõe reflexões a partir das questões levantadas à IPC na ocasião do XX encontro da ANPOF, sessão do GT de Filosofia das Ciências Físicas (§4).

---

[1] Raoni Arroyo é pesquisador pós-doutoral no Centro de Lógica, Epistemologia e História da Ciência (CLE), professor colaborador do Programa de Pós-Graduação em Filosofia da Universidade Estadual de Campinas (UNICAMP), e pesquisador do Grupo de Pesquisa em Lógica e Fundamentos da Ciência (CNPq). Seus principais interesses filosóficos são os aspectos metodológicos e epistemológicos da metafísica da ciência e sua relação com o realismo científico. E-mail: rwarroyo@unicamp.br.
[2] Pós-doutor PPGFIL/UFSC. Professor de Filosofia UAB/UFSC. E-mail: lmnf23@gmail.com.
[3] Pós-doutor CLE/Unicamp. Professor adjunto de Física e Filosofia no Centro de Ciência e Tecnologia em Energia e Sustentabilidade (CETENS), UFRB. E-mail: fredsantos@gmail.com.

## 2 A interpretação da consciência causal

O notório "problema da medição" representa um desafio significativo nos fundamentos da mecânica quântica, e não há consenso na literatura sobre se ele deve ser classificado como um problema científico ou filosófico (Albert, 1992). Grosso modo, o problema da medição se refere à dificuldade em explicar como o estado quântico, tendo como uma de suas características centrais o princípio da sobreposição, resulta em um estado definido após uma medição. Na mecânica quântica padrão, uma função de onda que pode representar uma variedade de resultados potenciais, mas ao representar diversos estados possíveis, o processo pelo qual um desses resultados se concretiza, na prática, ao observarmos o sistema, permanece controverso. Esse processo é chamado de "colapso", e questões acerca da sua causa têm sido matéria de debate por quase um século. A interpretação padrão assume que um estado definitivo surge apenas no momento da medição, sugerindo que o ato de observar tem uma participação direta na determinação da realidade do sistema. Esse fenômeno, no entanto, desafia explicações que buscam um entendimento claro do papel causal da observação no comportamento físico.

Levada às últimas consequências ontológicas, a mecânica quântica padrão afirma que o colapso ocorre por uma ação causal da consciência do observador (Wigner, 1983 [1961]). Em outras palavras, no momento em que uma mente consciente realiza a observação, a superposição de estados é substituída por um único estado objetivo. Essa abordagem tenta fornecer uma solução ao problema da medição ao situar a consciência como o catalisador de um evento físico mensurável, atribuindo ao observador um papel essencial na determinação do estado final. Seguindo Arroyo (2024, cap. 4), chamamos essa interpretação de "interpretação da consciência causal" (ICC).

A ICC, por sua vez, enfrenta uma séria crítica relacionada a uma questão clássica da metafísica/filosofia da mente, a saber, o dualismo mente-corpo. O dualismo tradicionalmente implica que a mente e o corpo são entidades fundamentalmente distintas, com a mente possuindo características não físicas que desafiam uma explicação causal nos termos da física. Para que a consciência afete o estado físico de um sistema quântico, a ICC pressupõe, em certa medida, que a mente exerce um tipo de causalidade que transcende o que se pode reduzir a processos físicos (para detalhes, ver Arroyo; Arenhart, 2019). Essa posição, que adota uma perspectiva dualista, desafia a integração entre o que é mental e o que é físico, levando a perguntas sobre a natureza da causalidade e a interação entre diferentes níveis de realidade. De acordo com o diagnóstico traçado em Arroyo (2024, cap. 5) e Arroyo; Nunes Filho; Moreira dos Santos (2024), a raiz do problema da ICC está em considerar a consciência dentro do *tipo ontológico*[4] de "substância", na qual o problema do dualismo emerge.

## 3 A interpretação da consciência processual

Nossa proposta defende que uma interpretação da consciência baseada na categoria ontológica de processos fornece alternativa à ICC no que diz respeito ao papel da consciência no processo do colapso. Crucialmente, ela contorna as dificuldades associadas ao dualismo mente-corpo enfrentadas pela ICC. Nesse cenário alternativo, todos os processos se manifestam de modo inter-relacionado, eliminando a necessidade de um agente consciente externo e independente para iniciar o colapso quântico. Chamemos essa interpretação de "interpretação da consciência processual" (ICP), cuja significância deve ser ressaltada em oposição ao uso da categoria de substância.

A categoria ontológica de processos é um elemento amplamente utilizado por Seibt (2009) em sua crítica ao substancialismo. De acordo com a filósofa a persistência da categoria de substância deve-se a certo apego histórico que temos à própria categoria.

> A popularidade da ontologia da substância, por exemplo, não se deve às suas (bastante pobres) realizações explicativas, mas principalmente ao fato de que o termo técnico substância é a categorização de um termo-gênero do raciocínio de senso comum que entendemos agentivamente particularmente bem: coisas. (Seibt, 2002, p. 60, tradução nossa)

Seibt (2002) define o conceito de um processo a partir da perspectiva de que a realidade não precisa ser interpretada como uma sala repleta de entidades particulares esperando passivamente para serem descritas. Na verdade, a cruzada de Seibt contra o conceito de substância impulsiona sua ontologia em direção a uma ontologia descentralizada do substancialismo e seu parente imediato, o particularismo. O principal desafio que ela enfrenta é deixar de lado o particularismo sem perder o individualismo. Além disso, ela precisa superar outras dificuldades, como definir um processo como uma categoria bem fundamentada. Assim, Seibt propõe uma ontologia de processo na qual não há espaço para o conceito de substância, nem para o de particular. Por sua vez, ela defende que os processos são indivíduos concretos contáveis ou incontáveis. Com isso em mente, precisamos entender como é possível construir indivíduos que não são particulares, e como os indivíduos podem ser contáveis ou incontáveis.

---

[4] Ver Arroyo e Arenhart (2024) para uma discussão sobre tipos ontológicos.

Em primeiro lugar, Seibt (2009) defende que a individualização não pode ser reduzida à particularização. Por particularização, pode-se entender duas coisas; uma entidade definida em termos de alguma alocação no espaço e no tempo; ou uma entidade definida em termos de uma relação sujeito-predicado com propriedades bem determinadas. Diferentemente dessa visão, ela sustenta que os indivíduos não precisam ser reduzidos nem à sua localização no espaço e no tempo nem à relação de dependência entre um sujeito e seus predicados (substancialismo). Ela argumenta que a concepção de que os indivíduos são individualizados por alguma *ecceidade* é incorreta.

Dessa forma, os particulares concretos são entidades semelhantes a coisas definidas principalmente na linguagem, por assim dizer, são entidades compactadas com limites e caracterizações bem conhecidas. Por outro lado, a ideia com processos é preservar a individualidade de vários tipos de entidades sem recorrer à particularidade. Por exemplo, um casamento é um processo contável individual que não é um particular como uma casa. Um casamento é algo que ocorre durante um intervalo de tempo em uma região do espaço, no entanto, não pode ser acumulado a ponto de ser considerado uma entidade única e pontualmente definida.

Assim como a ICP, a ICC é uma interpretação *ontológica* da mecânica quântica (em oposição às interpretações *epistêmicas*), segundo a qual os estados quânticos representam estados de coisas *no mundo*. A vantagem da ICP em relação à ICC é que ela permite um entendimento da consciência não como uma "substância" distinta, mas como uma dinâmica integrada ao próprio tecido da realidade quântica. Dessa forma, o ato de observação consciente e o colapso da função de onda seriam dois aspectos de um mesmo processo, coexistindo e interagindo sem exigir uma separação categórica entre mente e matéria. Em vez de recorrer ao dualismo, essa abordagem sustenta que o colapso da função de onda emerge naturalmente de interações que envolvem tanto o observador quanto o sistema observado, ambos compreendidos como partes de uma rede de processos em constante transformação.

Essa perspectiva processual oferece, portanto, uma solução mais parcimoniosa e ontologicamente unificada ao problema da medição, ao situar o papel da consciência dentro do fluxo da realidade física. Enquadrando a consciência como parte de uma estrutura dinâmica, nossa proposta reduz a necessidade de um elemento externo à mecânica quântica para explicar o colapso, unificando a causalidade em um esquema contínuo que abrange ambos os fenômenos quânticos e a experiência consciente. Historicamente, uma interpretação desse tipo foi idealizada/sugerida por Abner Shimony e Shimon Malin,[5] ainda que nenhum desses autores —nem qualquer outra pessoa, segundo nosso melhor conhecimento— levou a cabo a tarefa de especificar os detalhes de como seria uma interpretação desse tipo. Isso começou a ser feito em Arroyo (2024, cap. 5; ver também Arenhart, 2024) e Arroyo; Nunes Filho; Moreira dos Santos (2024).

---

[5] Shimony (1961), Shimony; Malin (2006); ver a lista de referências em Arroyo (2024) e Arroyo; Nunes Filho; Moreira dos Santos (2024).

## 4 Questões à IPC

A IPC foi apresentada para o GT de Filosofia das Ciências Físicas, na ocasião do XX encontro da ANPOF (2024), por Frederik Moreira dos Santos (Moreira dos Santos, Arroyo, Nunes Filho, 2024), e algumas questões foram levantadas. Nesta seção, as questões são abordadas coletivamente.

> **Questão 1:** Como se dá a relação entre o espaço e a atualização? O processo de atualização é a própria emergência do espaço? São sinônimos?
> **Questão 2:** Estamos trocando o papel da consciência pelo papel do espaço como causa, ou não há sentido em dizer que o espaço é o causador do colapso?
> **Questão 3:** Poderíamos descrever o surgimento dos padrões de interferência como uma relação entre as potencialidades e seus nexos? A interrelação entre propensidades com seus nexos criaria restrições para o espectro de possibilidades?

A ICP coloca como central a ideia de potencialidades e atualizações. De fato, na ICP, o problema da medição é traduzido para o problema da atualização das potencialidades. Se argumentássemos que a consciência é a *causa* da atualização das potencialidades, isso causaria (*sic*!), de imediato, uma tensão bastante conhecida: o retorno a uma ideia central da ICC, segundo a qual há um mecanismo pelo qual a consciência age causalmente. O problema dessa ideia, como apontado por Shimony e Hilary Putnam,[6] é que esse mecanismo deve ser especificado —e, no entanto, não o é. Por esse motivo, a ICP traça outra rota, afastando-se da ideia de causalidade descendente ao tratar a consciência como um dos pólos das entidades processuais, interdependente do pólo físico, no *continuum* da experiência.

Visando escapar aos tradicionais dualismos ligados à dicotomia mente-corpo ou percepção e percebido, a ICP sustenta que a assim chamada "consciência" não é um pólo separado ou superior ao ato de percepção. A consciência não se opõe a qualquer entidade, sendo, na verdade, um polo que emerge da própria atualização das entidades. Nesse sentido, a consciência não está sendo substituída, mas relida sobre uma nova perspectiva teórica alinhada à teoria de processos[7].

---

[6] Para um estudo detalhado da controvérsia, ver Moreira dos Santos; Pessoa Jr. (2011).
[7] Para o "novo" *locus classicus* da concepção fenomenológica no contexto quântico, ver French (2023); para um contraponto crítico, ver Arroyo; Nunes Filho (2018).

A ICP soluciona, portanto, o problema da medição por outra via. Reminiscente da filosofia processual de Alfred N. Whitehead (1928), potencialidades não possuem existência espaço-temporal, ao passo que entidades atuais —que constituem a base ontológica da filosofia whiteheadiana— se atualizam por atos de autocriação. De acordo com a ICP, quanto maior for o grau de liberdade do sistema em questão, maior a tendência desse sistema —outrora meramente potencial/superposto— colapsar/autocriar-se enquanto uma entidade atual. Essa característica faz com que a ICP se assemelhe às teorias de colapso espontâneo,[8] o que a afasta da mecânica quântica padrão.

Assim, pode-se dizer que o colapso, entendido como o ato de autocriação de entidades atuais, o que significa uma transição entre o potencial (não espaço-temporal) para a espaço-temporalidade, é análogo à emergência do espaço-tempo. De fato, ao considerar os graus de liberdade de um sistema, o que estamos assumindo, a depender da formulação —espaço de configuração ou espaço de fases— são as potencialidades do sistema, e não aquilo que é obtido no momento na medição. Formalmente, podemos tratar da medição apenas ao considerar a estrutura matemática (em geral uma variedade riemanniana) na qual o espaço e o tempo são caracterizados em dada teoria. A estrutura do espaço-temporal é o que caracteriza a atualização do dado dos fenômenos (ver Castellani, 1998).

Isso, no entanto, traz consigo uma séria dificuldade, não antecipada por nós em Arroyo (2024) e Arroyo; Nunes Filho; Moreira dos Santos (2024), que diz respeito à indeterminação metafísica quântica (*quantum metaphysical indeterminacy*).[9] Suponha que eventos não-colapsados sejam meramente potenciais. Como parte da interpretação tardia de Heisenberg (1958), estão em "*potentia*". Como é bem conhecido, Heisenberg, em seus escritos tardios, introduz, nas palavras de Shimony (183, p. 212, tradução nossa) "[...] uma modalidade que se encontra entre a possibilidade lógica e a atualidade, que ele chama de 'potentia'". Como parte da interpretação whiteheadiana, não são entidades espaço-temporais; em suas próprias palavras: "as entidades atuais [...] tornam real o que anteriormente era apenas potencial" (Whitehead, 1928, p. 72, tradução nossa). A ICP une ambos os aspectos. O problema disso está em explicar fenômenos de interferência. Notavelmente, como se observa em diversos fenômenos quânticos, como o experimento da fenda dupla (ver Pessoa Jr., 2019, cap. 1), franjas de interferência são observadas nos anteparos de detecção após uma série de detecções, mesmo havendo apenas uma entidade quântica (*e.g.*, um elétron) de cada vez no aparato.

A seguinte questão segue-se naturalmente: *por qual das fendas cada elétron passou*

---

[8] Ver Allori *et al*. (2021) para perspectivas contemporâneas sobre esse programa de pesquisa nos fundamentos da mecânica quântica.
[9] Para uma discussão panorâmica sobre esse assunto, e uma lista com as referências fundamentais sobre o debate, ver Calosi e Mariani (2021).

*individualmente?* Nossas quatro inferências básicas são falsas. É falso supor que o elétron passou por uma das fendas, por outra, por ambas, e por nenhuma delas (Albert, 1992). Diz-se então que o elétron passa *pela superposição das duas fendas*. Atribuir sentido físico a essa inferência ocupa grande parte da tarefa interpretativa nos fundamentos da mecânica quântica. No entanto, se entendermos, como a ICP parece sugerir, que a superposição é um evento fora do espaço-tempo, a interpretação carrega consigo o ônus de explicar como o fenômeno de interferência ocorre. De outro modo, é uma interpretação empiricamente inadequada.

Esse é um problema análogo às interpretações do tipo "*sparse view*" ou "*gappy*" acerca da indeterminação metafísica quântica, já que devemos supor que o elétron está *em algum lugar do espaço-tempo* para explicar o fenômeno de interferência. A *sparse view* é a interpretação segundo a qual, em situações de indeterminação (como de superposição, no caso de questões do tipo "qual fenda?") os objetos simplesmente não possuem a propriedade. Nesse caso, de localização. Não é que os objetos não possuam o valor da propriedade; o caso é que eles não possuem *a propriedade* em si (*e.g.*, não possuem posição). Já a abordagem "*gappy*" da indeterminação metafísica quântica diz que os objetos não possuem o valor (nesse caso, possui posição mas nenhum *valor* de posição). Ambas as interpretações enfrentam dificuldades de explicar o fenômeno —bem documentado— da interferência, e por isso são severamente criticadas. Ao que parece, a ICP deve ser criticada sob os mesmos critérios, caso queira-se manter uma visão ontológica dos estados quânticos.

Uma possível resposta está em considerar que a interação entre potencialidades poderia ter efeitos espaço-temporais, sem que, no entanto, sejam atualizadas. Isso requer uma modificação na ICP no que diz respeito à estrita não-espaço-temporalidade do domínio ontológico das potencialidades. Nessa versão modificada, pode-se entender que uma entidade atual é, não apenas espaço-temporal, mas *determinada*, ao passo que a indeterminação metafísica, característica da superposição, possa ser entendida em termos "*glutty*", isto é, instanciando *mais de um valor* da propriedade (no caso, localização). Isso nos levaria fundo na metafísica da mereologia de multilocalização (Kleinschmidt, 2011; Calosi, 2022), sendo um campo aberto para futuras investigações da ICP. A relação entre teoria de processos e mereologia apresenta-se muito promissora, pois coaduna com vários aspectos formais do domínio quântico (Seibt, 2002). Por hora, essa indicação deve ser suficiente.

## 5 Considerações finais

Notavelmente, as respostas a essas questões nos levam além daquilo que foi desenvolvido nos primórdios da ICP (Arroyo, 2024; Arroyo; Nunes Filho; Moreira dos

Santos, 2024) — o que é de se esperar, já que trata-se de uma interpretação (ou programa de pesquisa) em construção.

## Agradecimentos

## Financiamento

## Declaração de Contribuição dos Autores

As contribuições dos autores neste trabalho foram descritas com base na taxonomia CRediT (*Contributor Roles Taxonomy*), conforme detalhado a seguir:
- Raoni Arroyo: Conceituação, Metodologia, Investigação, Redação – Rascunho Original.
- Lauro de Matos Nunes Filho: Investigação, Redação – Rascunho Original.
- Frederik Moreira dos Santos: Redação – Revisão e Edição.

## Referências

ALBERT, D. Z. *Quantum mechanics and experience*. Cambridge: Harvard University Press, 1992.

ALLORI, V.; BASSI, A.; DÜRR, D.; ZANGHI, N. (eds.). *Do Wave Functions Jump? Perspectives on the Work of GianCarlo Ghirardi*. Cham: Springer, 2021.

ARENHART, J. R. B. Resenha de "Consciência e mecânica quântica: uma abordagem filosófica". *Princípios*, v. 31, n. 66, 2024. DOI: 10.21680/1983-2109.2024v31n66ID37248.

ARROYO, R.; NUNES FILHO, L. d. M. "Underdeterminations of Consciousness in Quantum Mechanics". *Principia*, v. 22, n. 2, p. 321–337, 2018. DOI:

10.5007/1808-1711.2018v22n2p321.

ARROYO, R. *Consciência E Mecânica Quântica: Uma Abordagem Filosófica*. São Paulo: LF Editorial, 2024.

ARROYO, R.; ARENHART, J. R. B. Between physics and metaphysics: a discussion of the status of mind in quantum mechanics. *In*: DE BARROS, J. A.; MONTEMAYOR, C. (ed.). *Quanta and mind: essays on the connection between quantum mechanics and consciousness*. Cham: Springer, 2019. p. 31–42.

ARROYO, R.; ARENHART, J. R. B. Quantum ontology de-naturalized: what we can't learn from quantum mechanics. *Theoria. An International Journal for Theory, History and Foundations of Science*, v. 39, n. 2, p. 193–218, 2024.

ARROYO, R., NUNES FILHO, L. d. M.; MOREIRA DOS SANTOS, F. "Towards a Process-Based Approach to Consciousness and Collapse in Quantum Mechanics", *Manuscrito*, v. 44, n. 1, p. e-2023-0047-R1, 2024.

CALOSI, C. There Are No Saints, Or: Quantum Multilocation. *Grazer Philosophische Studien*, v. 99, n. 1, p. 30–49, 2022.

CALOSI, C.; MARIANI, C. "Quantum indeterminacy". *Philosophy Compass*, v. 14, n. 4, p. e12731, 2021.

CASTELLANI, E.. "Galilean Particles: An example of constitution of objects". *In*: CASTELLANI, E. (ed.). Interpreting bodies: classical and quantum objects in modern physics. Princeton: Princeton University Press, 1998.

FRENCH, S. *A Phenomenological Approach to Quantum Qechanics: Cutting the Chain of Correlations*. Oxford: Oxford University Press, 2023.

HEISENBERG, W. *Physics and Philosophy: The Revolution in Modern Science*, Nova Iorque: Harper e Row, 1958.

KLEINSCHMIDT, S. "Multilocation and Mereology". *Philosophical Perspectives*, v. 25, p. 253–276, 2011.

MOREIRA DOS SANTOS, F.; PESSOA JR., O. "Delineando o problema da medição na mecânica quântica: o debate de Margenau e Wigner versus Putnam". *Scientiæ Studia*, v. 9, n. 3, p. 625-644, 2011.

WHITEHEAD, A. N. *Process And Reality: An Essay In Cosmology*. Nova Iorque: Free Press, 1928.

PESSOA JR., O. *Conceitos de Física Quântica*, Volume I. 4ª ed. São Paulo: LF Editorial, 2019.

SEIBT, J. "Forms of emergent interaction in General Process Theory". *Synthese*, v. 166, p. 479–512, 2009.

SEIBT, J. "Quanta, Tropes, or Processes: Ontologies for QFT Beyond the Myth of Substance". *In*: *Ontological aspects of quantum field theory*. New Jersey, London,

Singapore, Hong Kong: World Scientific, 2002. p. 53–97.

SHIMONY, A. "Shimony to Wigner on May 31, 1961". *In*: Eugene Paul Wigner Papers, Collection C0742, Box 83, Folder 7, Manuscripts Division, Department of Special Collections. Princeton: Princeton University Library, 1961.

SHIMONY, A. Reflections on the philosophy of Bohr, Heisenberg, and Schrodinger. *In*: COHEN, R. S.; LAUDAN, L. (eds.). *Physics, philosophy and psychoanalysis: essays in honor of Adolf Grunbaum*. Dordrecht: D. Reidel, 1983. p. 209–221.

SHIMONY, A.; MALIN, S. "Dialogue Abner Shimony–Shimon Malin". *Quantum Information Processing*, v. 5, n. 4, p. 261–276, 2006.

WIGNER, E. "Remarks On The Mind-Body Question". *In*: WHEELER, J. A.; ZUREK, W. H. (eds.). *Quantum Theory and Measurement*. Princeton: Princeton University Press, p. 168–181, 1983 (reimpr. de "Remarks on the Mind-Body Question", em The Scientist Speculates, ed. por GOOD, I. J., Heineman, 1961).